\documentclass[12pt,preprint]{aastex}

\shorttitle{Impostor }
\shortauthors{Humphreys et al. }

\begin{document}

\title{Multiple Outflows in the Giant Eruption of a Massive Star\altaffilmark{1} }

\author{
Roberta M.. Humphreys\altaffilmark{2}, 
John C. Martin\altaffilmark{3}, 
Michael S. Gordon\altaffilmark{2},  
and 
Terry J. Jones\altaffilmark{2}
}

\altaffiltext{1}  
{Based on observations  obtained with the Large Binocular Telescope (LBT), an international collaboration among institutions in the United
 States, Italy and Germany. LBT Corporation partners are: The University of
 Arizona on behalf of the Arizona university system; Istituto Nazionale di
 Astrofisica, Italy; LBT Beteiligungsgesellschaft, Germany, representing the
 Max-Planck Society, the Astrophysical Institute Potsdam, and Heidelberg
 University; The Ohio State University, and The Research Corporation, on
 behalf of The University of Notre Dame, University of Minnesota and
 University of Virginia.  } 

\altaffiltext{2}
{Minnesota Institute for Astrophysics, 116 Church St SE, University of Minnesota
, Minneapolis, MN 55455; roberta@umn.edu} 

\altaffiltext{3}
{University of Illinois Springfield, Springfield, IL 62703}

\begin{abstract}
The supernova impostor PSN J09132750+7627410 in NGC 2748 reached a maximum
luminosity of $\approx$ -14 mag. It was quickly realized that its was not
a true supernova, but another example of a non-terminal giant eruption. 
PSN J09132750+7627410 is distinguished by multiple P Cygni absorption 
minima in the Balmer emission lines that correspond to outflow velocities
of -400, -1100, and -1600 km s$^{-1}$.  Multiple outflows have been observed
in only a few other objects. In this paper we describe the evolution of the 
spectrum and the P Cygni profiles for three months past maximum, the 
post-maximum formation of a cool, dense wind, and the identification
of a possible progenitor. One of the possible progenitors is an infrared source. Its pre-eruption spectral energy distribution suggests a bolometric luminosity of -8.3 mag and a dust temperature of 780\arcdeg K. If it is the progenitor it is above the AGB limit unlike the intermediate luminosity red transients.  The three P Cygni profiles could be due to ejecta from 
the current eruption, the wind of the progenitor, or previous mass loss
events. We suggest that they were all formed as part of the same high mass loss
event and are due
to material ejected at different velocities or energies. We also suggest that
multiple outflows during giant eruptions may be more common than reported. 
\end{abstract} 


\section{Introduction}
Transient surveys  are finding an increasing number of 
what appear to be non-terminal giant eruptions.  Many of these giant 
eruptions are spectroscopically similar to Type IIn
supernovae and thus receive a supernova (SN) designation, but are later recognized as
sub-luminous or their spectra and light curves do not develop like true
supernovae. Consequently, they are often referred to as ``supernova impostors''
\citep{VanDyk2000}. These impostors or giant eruptions are examples of high
mass loss episodes apparently from evolved massive stars (see Van Dyk \& Matheson 2012 for a review and references therein). Authors often refer 
to them  as Luminous Blue Variables (LBVs), but these giant eruptions are 
distinctly different from  LBV/S Doradus variability in which the 
star does not increase in luminosity and the eruption or maximum  light can last for several years. 

PSN J09132750+7627410 did not receive a supernova designation. 
It was posted  on the CBAT Transient Objects Confirmation Page on 2015 February 10 by K. Itagaki as a possible supernova (PSN) in NGC 2748. The first reported
magnitudes of 17.7 to 18.1 suggested a luminosity of $\approx$ -14 mag based
on membership in NGC 2748 at a mean distance of 20.97 Mpc via NED, and based on 
Tully-Fisher and kinematic distances. Thus, it  
was likely not a true supernova, but another example of a giant eruption. 
A spectrum obtained only a day later, on 2015 February 11 reported  by \citet{Tartaglia}  
showed narrow emission lines. They also measured an apparent V magnitude of 18.7 and a luminosity of -13 mag. Based on the  spectral appearance and the 
luminosity, they suggested that this PSN was  a SN impostor.

Our first spectrum of PSN J09132750+7627410 \citep{RH2015} observed on 
2015 February 16, showed multiple P Cygni absorption components in the   
prominent Balmer emission lines.
These features could be due to ejecta
from the current eruption, from previous mass loss events, or the wind
of the progenitor. Evidence for multiple outflows have been previously
observed  in the spectra of only a few objects; two impostor eruptions (SN2000ch, SN Hunt 248), the peculiar SN2009ip, and a 
Type IIn supernova (SN2005gj), see \S {4}. For that reason we obtained 
additional spectra and photometry. In this brief {\it Paper} we present 
our observations, describe the spectrum and our measurements, discuss  
 a possible progenitor, and the origin of the multiple P Cygni absorption features.

\section{Observations} 

\subsection{Spectroscopy}

Moderate resolution spectra of PSN J09132750+7627410 were observed with the MODS1 spectrograph 
on the Large Binocular Telescope (LBT) in February, April and May, 2015.
The MODS1 uses a dichroic to obtain blue and red spectra simultaneously 
with the  G400L and G750L gratings, 
respectively. The total wavelength coverage is from 3200{\AA}  to 
more than 1$\mu$m.  We used a 1\arcsec slit yielding a resolution of  1500 in the blue and  2000 in the
red. This gives a velocity resolution of 150 km s$^{-1}$  at H$\alpha$. The two dimensional spectra were initially reduced using  
the {\it modsCCDred} pipeline for bias subtraction and flat fielding. The spectra 
were then extracted, and  wavelength and flux calibrated using 
the standard  IRAF twodspec and onedspec packages. The extracted spectra cover the wavelength ranges  3600 -- 5600 and 6000 -- 9000 {\AA}. The MODS1 uses standard lamps exposed 
during the afternoon for wavelength calibration, but despite a flexure model, there
remain wavelength calibration uncertainties. For this reason we used the night sky lines
for the wavelength calibration in the red. This may introduce a small velocity offset between the
blue and red spectra. But, for this object in NGC 2748 (velocity +1476 km sec$^{-1}$), the
difference is not significant for our discussion.

The journal of observations is given in Table 1. 

\subsection{Photometry During the Eruption}  

Broadband CCD photometry with ASTRODON Johnson-Cousins filters was obtained with
 the 20-inch telescope at the Barber Observatory with an Apogee U42 CCD camera 
 using a back-illuminated E2V CCD42-40 chip.  All images were flat-fielded and 
 bias and dark subtracted.  Aperture photometry was measured using VPHOT (https://www.aavso.org/vphot) developed primarily by Geir Klingenberg (coding and 
 design) and Arne Henden (photometry).  Photometry was measured in circular 
 apertures with a diameter 3.0 times the FWHM of the stellar profiles.  
 Sky background was sampled from an annulus centered on each star with an inner radius of 11.5$\arcsec$ and 3$\arcsec$ wide.  No effort was made to subtract 
 the underlying contribution from the galaxy.  The galaxy contribution is 
 estimated from images without the target present to be  
 V $=$ 21 mag arcsec$^{-2}$ (about V= 19 mag in a typical aperture).  
    
The brightness of the target is an unweighted average  measured with respect to 
 9 reference stars in the magnitude  range  V = 14.0 to 16.2 and within 10 arcminutes 
of the galaxy (but well separated from it) selected from the UCAC4 catalog 
(Zacharias et al. 2013) using the prescription of Toone (2005).  
Photometric error for the target was calculated as the quadrature sum of 
the CCD equation noise for the target, plus the standard deviation of the 
reference stars.  We used the   standard photometric transformations for 
the optical system. 

The photometric observations are summarized in Table 2 which also includes the 
 discovery magnitudes from the CBAT “Transient Objects Followup Reports”. 
The earlest photometry observed during the first 24hrs shows a decline of  
0.6 mag which could be due to a difference in calibration. We note though
that the early I~band magnitude is consistent with later photometry.
 Additional post-maximum magnitudes were recorded in the Pan-STARRS Survey 
 for Transients. The target is PS15jf. The r(P1) and i(P1) magnitudes in Table 2 are on the Pan-STARRS filter system (Tonry et al. 2012).

Although the photometric record is sparse, and the onset of the eruption is uncertain, the available data suggest that this eruption was 
relatively brief. PSN J09132750+7627410 was at or near maximum  light for 
only about a month. The photometry shows that a decline had begun by about 30 days post-maximum. Interestingly, our data point from  2015 May 19 suggests that the object may have  brightened somewhat. This is supported by the 
flux-calibrated spectrum from May 20 observed under good conditions. Short-term oscillations in brightness are common in these objects as is evident from the erratic
behavior of SN2009ip \citep{Pastorello12,Margutti,Martin}. To check on variability and a 
possible recovery, we observed it again in 2016 January and February. 
The target remained below our limit of detectability  defined as the median value in the aperture that was not statistically different  from the median value in the background annulus.  We note that the seeing
and observing conditions were excellent for the February measurement.

\subsection{Pre-Eruption Images and a Possible Progenitor}

NGC 2748 is a well-studied spiral galaxy at a distance of 20.97 Mpc, modulus 31.60 mag, from 
NED.
It is the site of a super-massive black hole in its nucleus and two previous supernovae, 
SN1985A (Ia) and SN2013ff (Ic). Consequently, there are numerous space-based images in the 
Mikulski Archive for Space Telescopes (MAST), but PSN J09132750+7627410 is in the outer parts
of the galaxy, and was just off the frame in most of the {\it HST} images. 
We identified the target field on four {\it HST}/WFPC2 images of NGC 2748; 
two with F450W and two with F814W. (GO-9042,PI:Smartt) obtained 2001 July 6. 
The exposure times were 230s each and the target was on the WF3 chip. 
We processed the images using DOLPHOT \citep{Dolphin} which successfully 
fit a PSF to six objects within 2\arcsec of the target position with sigma
threshold 2.5, the default. We only considered DOLPHOT detections consistent with a point source (class 1 and 2) using the default WFPC2 DOLPHOT settings. 
Only two objects were above the 
sky background in all four images. The resulting photometry in VEGAMAGs 
for these two 
objects is in Table 3. The F450W and F814W filters are  comparable to the Johnson B and Cousins I bands, respectively, yielding an instrumental F450W -F814W  or {\it b-i} color.  

For comparison with the 
target's published position, which we confirmed with astrometry from our CCD images\footnote{We measured an average postion of  RA 09:13:27.55,	Dec +76:27:41.1	in ICRS	J2000 coordinates with a standard deviation of 0$\farcs$2 from six separate images using GSC 2.3 stars in the field with  astrometry.net \citep{Lang}. This agrees very well with the discovery position.}, we determined the offsets of their WCS image 
centers  
relative to ICRS J2000 coordinates using four GSC 2.3 stars  on the same frame,  shifted in X and Y and with rotation with a  standard deviation
of 0$\farcs$4 in RA and Dec. The combined uncertainty of the target's position and of the WCS in the HST image is then 0$\farcs$45.  Their ICRS positions are included in Table 3. The two summed  HST frames are shown in Figure 1, centered on the impostor's position and corrected for the 
offset. Star A is  within the small circle, radius 0$\farcs$2, 
centered on the target's position.  The fainter star
B, further from the center, is just outside a circle with the 0$\farcs$45 radius. It is apparently quite red, but  with a large uncertainty. Based on its position,  star A is the more 
likely progenitor. Its absolute blue magnitude (F450W) would be $\approx$ -7.3. with a foreground
 galactic extinction A$_{B}$ of 0.097 \citep{Schlafly2011}. Its extinction-corrected {\it b-i} color of 0.58 mag implies an A-type or early F-type supergiant.  
 NGC 2748 has prominent dust lanes, although none are apparent close to the target, so there may be some  
additional internal extinction and star A would be both brighter 
and bluer. There is also an extended area of low level emission around 
star A visible in the F814W image which could be due to unresolved stars. 

\begin{figure}[ht!]  
\figurenum{1}
\epsscale{0.8}
\plotone{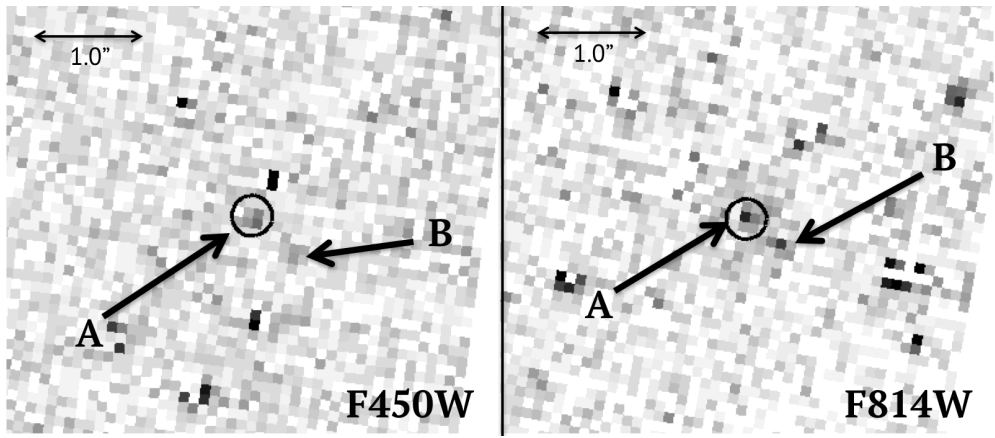}
\caption{The summed HST/WFPC2 F450W and F814W images including the region
around PSN J09132750+7627410. 
The smaller dashed circle 0$\farcs$2 in radius is centered on the position
of the impostor.  The cosmic rays have not been removed. }
\end{figure}

We have also identified  a  source in the  Spitzer Science Archive 
in three separate epochs at 3.6$\mu$m and 4.5$\mu$m   at nearly 
  the same position. Since the background is high and variable, we
  initially used a profile fitting routine to determine the flux distribution
  across the image and the appropriate radius for aperture photometry.   
  We then used the Astropy-affiliated package ``photutils'' 
  to measure aperture photometry on the Level 2 PBCD mosaic images.  
  Since there is a significant gradient in the infrared background, 
  particularly in the 4.5\micron\ images, the sky is modeled as a 
  two-dimensional polynomial surface across the source and subtracted 
  from a 1.8\arcsec radius aperture.  In each image, subpixel centroids 
  are calculated as the center of mass determined from image moments.  
    The  magnitudes 
     are included in  Table 3 with errors calculated from the uncertai
     nty maps provided by the Spitzer Science Archive for each field.  
The IRAC images from 2014 have the best signal to noise and are shown
  with the HST/WFPC2 F814W image for comparison in Figure 2. 
  We found   no systematic  offset between the the IRAC position and   
  the ICRS J2000 using the same reference stars, but with a standard 
  deviation of
  0$\farcs$64 and 0$\farcs$46 in RA and Dec respectively.  The
  WFPC2 and IRAC frames are thus on the same coordinate system. 
 The IRAC position is included in Table 3. 
We were not able to identify the source in the WISE survey, because it is
  not resolvable from the background galaxy.

\begin{figure}[h!]  
\figurenum{2}  
\epsscale{0.8}
\plotone{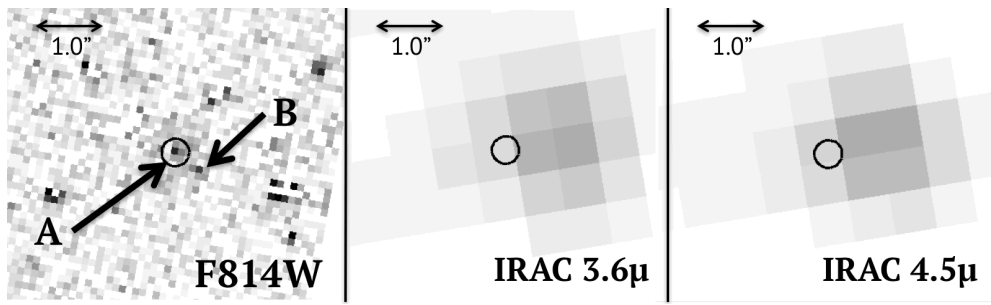}
\caption{The HST/WFPC2 F814W image and IRAC 3.6$\mu$m  and 4.5$\mu$m images
 from 2014 of the region around PSN J09132750+7627410.  The WCS coordinates 
 were corrected to the ICRS J2000 using the same reference stars so that the
 WFPC2 and IRAC frames are on the same coordinate system..}
\end{figure}

The spectral energy distributions (SED) are shown in Figure 3 for stars  
A and B and  the IRAC source. 
   The IRAC source is very red with a        
  rising energy distribution to longer wavelengths  
  similar to obscured AGB stars and OH/IR stars \citep{Jones}  with extensive circumstellar dust.  Without  longer 
  wavelength data it is not possible to know where the SED peaks. We show 
  Planck curves fits to the data points from  2014 with a color temperature of 
  780\arcdeg K. The source could be both cooler and more luminous. Planck curves are also shown  fit  to the optical photometry for stars A and B.

\begin{figure}[h!]  
\figurenum{3}
\epsscale{1.0}
\plotone{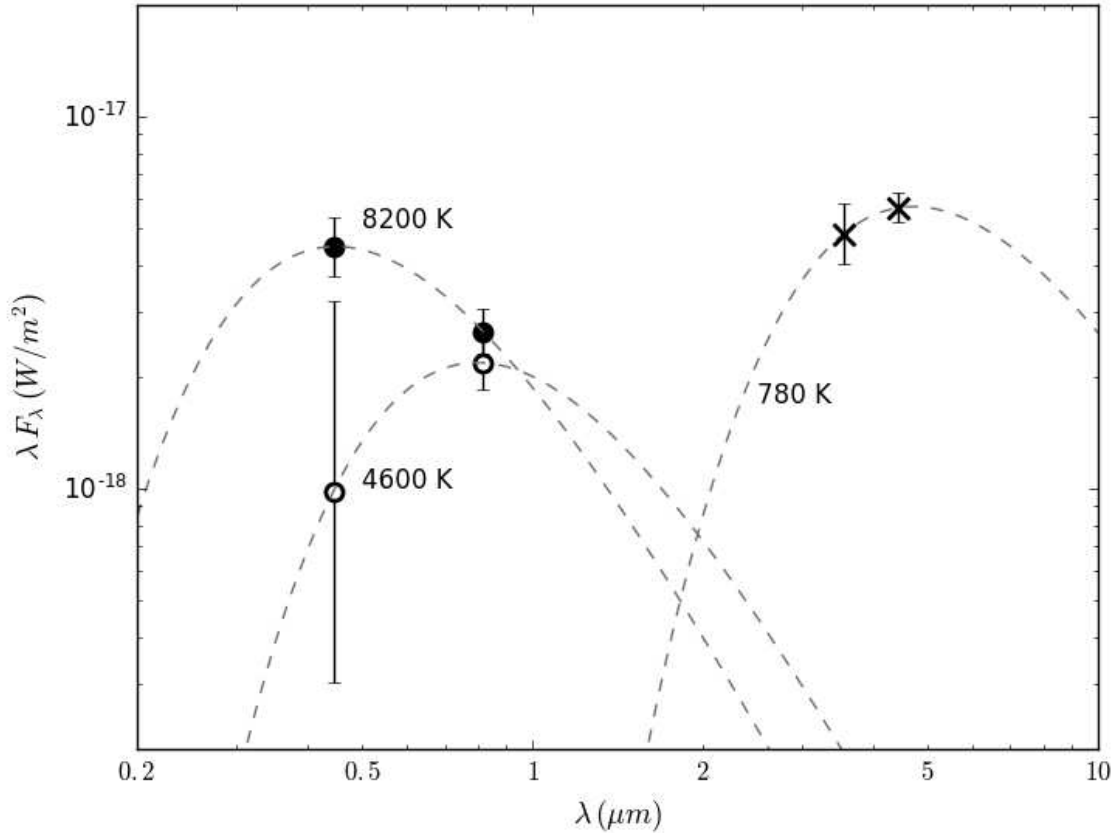}
\caption{The SEDs  for stars A and B and the infrared source near the position 
of PSN J09132750+7627410.  Planck curve  fits are shown with color temperatures 
of 8200\arcdeg K  and 4600\arcdeg K  for stars A , filled circles, and B, 
open circles,  and 780\arcdeg K  for the infrared source. No corrections for possible interstellar extinction have been applied.}
\end{figure}

We have two possibilities for an optical progenitor; star A or B, alone or
in combination with the infrared source.  The  photometry for star A is incompatible with the IR SED.  If A is the progenitor, it would be an
intermediate-type supergiant of relatively low luminosity (9 $\times$ 10$^{4}$ L$_{\odot}$, M$_{Bol} = -7.6$ mag) and initial mass $\approx$ 15 -- 20 M$_{\odot}$.
Star B  has a luminosity of 4.4 $\times$ 10$^{4}$ L$_{\odot}$ and based on its 
somewhat uncertain color, it  could be an evolved star of approximately 12 -- 15 M$_{\odot}$.   Based on these limited SEDs, it is possible that the infrared source could be associated with  star B.  If so, then star B's luminosity would
be dominated by the infrared radiation.

Given the lower spatial resolution with IRAC, another possibility is that the infrared source,  possibly with no optical 
counterpart, is the progenitor. The postions of the impostor and the IRAC source marginally agree at the 1 sigma level and are coincident within two sigma of their
positional uncertainties. If this is the case, PSN J09132750+7627410 could be similar to the dusty  
Intermediate Luminosity Red Transients (ILRTs) like SN 2008S \citep{Smith09} and the 2008 NGC 300 OT \citep{Bond09,RMH2011}. Assuming that the infrared source is similar to an AGB, we use its 3.6$\mu$m  $-$ 4.5$\mu$m  color index in 2014 
with the bolometric calibration from
 \citet{Blommaert} to derive a bolometric magnitude of -8.3  mag (1.6 $\times$ 10$^{5}$ L$_{\odot}$).  Integrating the Planck curve gives a luminosity of 
 10$^{5}$ L$_{\odot}$.
 This luminosity is well above the nominal AGB limit at M$_{Bol} \cong$ -7.0 mag. The
 object is  more luminous and therefore more massive that the ILRTs which are
 close to the AGB limit, see the HR diagram (Fig. 14)  in \citet{RMH2011}. 
 It is too red to be a foreground dwarf. 
 Alternatively, it could be an H~II region or the chance superposition
  of a background object such as an AGN, but we could not find any catalogued
   objects at its position.
 
 Assuming that the infrared source is in NGC 2748, with this  luminosity, 
 the candidate would of course be a massive star of $\approx$ 25 M$_{\odot}$. 
 It could be a very 
 dusty red supergiant perhaps similar to the OH/IR stars in our galaxy. 
 They are potential supergiants,  but their distances are not known.  \citet{Thompson} and \citet{Khan2015} have identified several luminous, optically obscured  stars in 
 M33 and other nearby galaxies which may be similar to the supergiant OH/IR stars.  The infrared source's mid-infrared color and dust temperature are similar to many of the  sources listed by \citet{Khan2015}, although their objects are 
 significantly more luminous with  log L $\sim$ 5.5 to 6.0 L$_{\odot}$.  

It is tempting to identify the infrared source with the
 erupting star, but more information was needed. Time was requested with 
 Spitzer to confirm  if the IRAC source had survived or changed after the 
 eruption, but was denied.  Based on their positions, stars A and B and the
 infrared source are all possible progenitors, but we favor a tentative 
 identification with the infrared source.  

If the infrared source is the progenitor, the underlying star is most likely
a cool or intermediate temperature evolved supergiant that may be 
transiting the HR Diagram to the blue, similar to  more luminous examples
such as VarA in M33. In this transition, the stars enter a period of enhanced
instability that leads to high mass loss episodes. Followup imaging 
of PSN J09132750+7627410 with HST and Spitzer would be very worthwhile.

\section{The Spectrum: Multiple P Cygni Absorptions} 

Our first and highest quality spectra from  2015 February 16   
were observed about six days after the reported discovery and the presumed maximum. The blue spectrum is shown in Figure 4.  
The blue and red spectra show strong narrow Balmer 
emission lines with prominent P Cygni features from  H$\alpha$ to H$\epsilon$.  
Three absorption 
minima are present in H$\beta$, H$\gamma$, H$\delta$ and at least two are clearly identified  in
H$\alpha$. The H$\alpha$ and H$\beta$ profiles are shown
 in Figure 5 with the mulitple absorption minima identified.  
  In addition to the narrow peaks, the hydrogen emission 
 profiles all show the classic asymmetric Thomson scattering 
profile with prominent red wings extending to more than 2000 km sec$^{-1}$ at H$\alpha$ and H$\beta$ due to scattering off the electrons in the wind not Doppler motion.  Because of the strong scattering wings, we measured the widths
of the H$\alpha$ and H$\beta$ lines above where the profile begins to broaden.
The ``FWHM'' of the H$\alpha$ and H$\beta$ narrow peaks are 379 km  s$^{-1}$ and 368 km  s$^{-1}$, respectively\footnote{The FWHM was measured at 1.62 $\times$ 10$^{-16}$ ergs s$^{-1}$ cm$^{-2}$ {\AA}$^{-1}$ for H$\alpha$ and at 
1.25 $^{-16}$ ergs s$^{-1}$ cm$^{-2}$ {\AA}$^{-1}$ for H$\beta$}. The narrow Balmer emission peaks have  a mean velocity of
1685 km  s$^{-1}$,  and the individual lines show no significant velocity shift in
the two later spectra.  Although, this is about 200 km  s$^{-1}$ greater than the
published Doppler velocity for NGC 2748 of +1476 km  s$^{-1}$, the object
is in the outermost parts of the galaxy, and the velocity difference is consistent with rotation at its distance from the center \citep{Atkinson}.

\begin{figure}[h!]   
\figurenum{4}
\epsscale{1.0}
\plotone{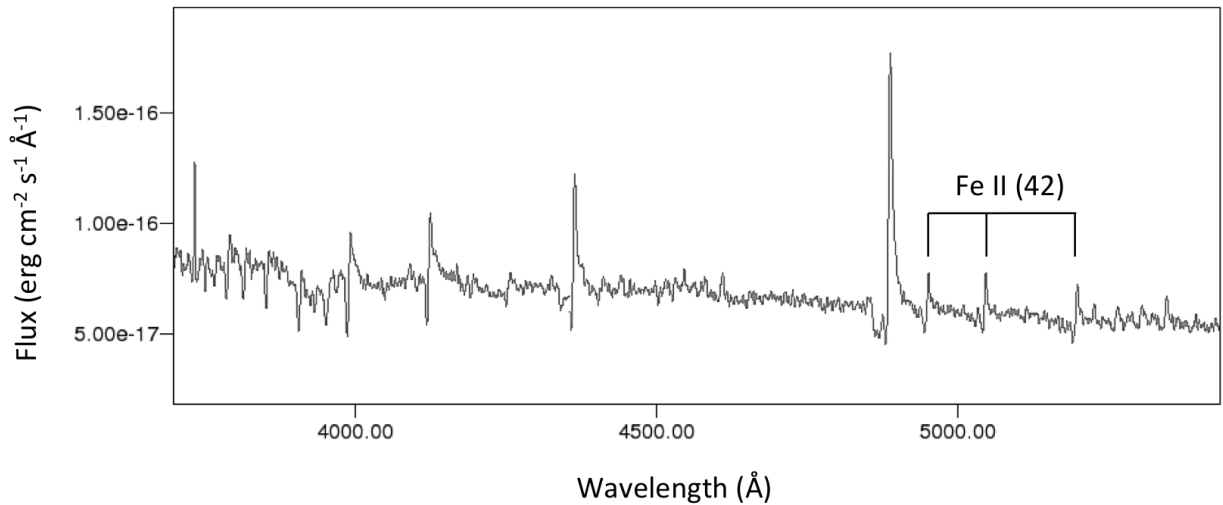}
\caption{Blue spectrum from  day 6}
\end{figure}

\begin{figure}[h!]   
\figurenum{5}
\epsscale{1.0}
\plotone{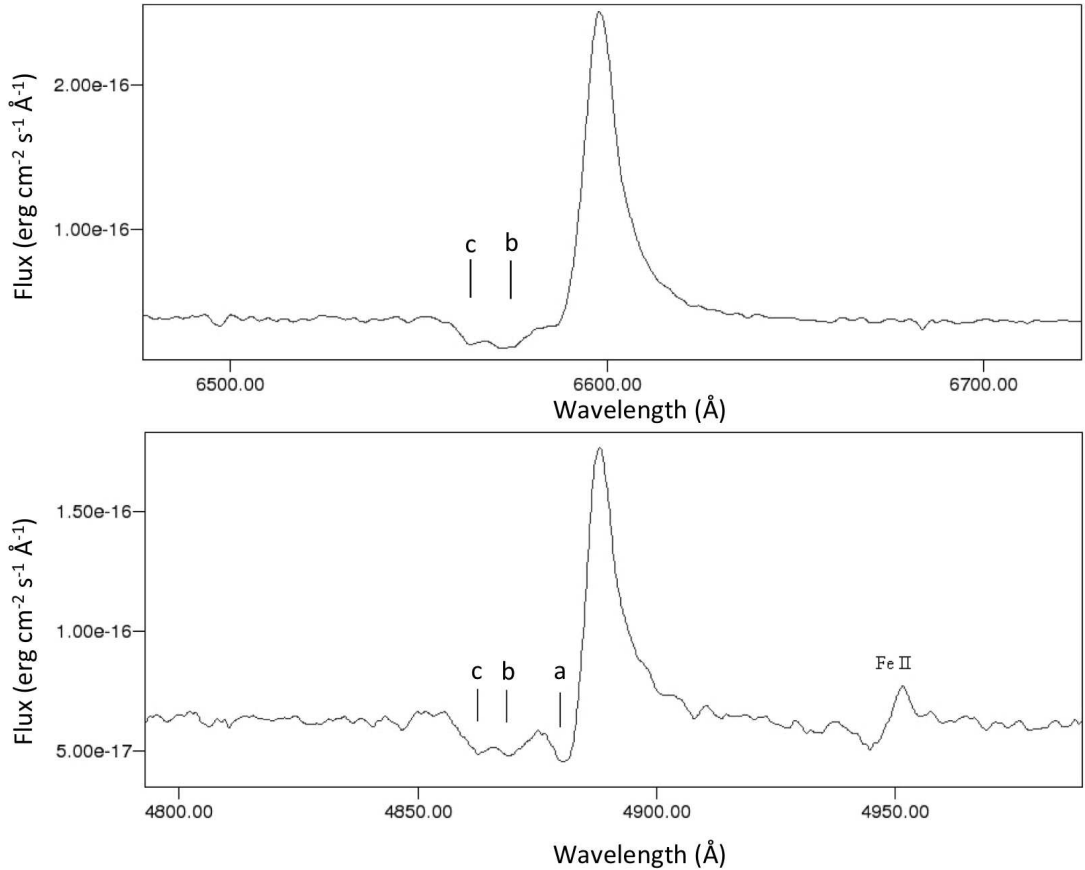}
\caption{ The profiles of the H$\alpha$ (top) and H$\beta$ (bottom) lines from  
Day 6 showing the multiple P Cygni absorption features.}
\end{figure}

 The higher Hydrogen lines at $\lambda$$\lambda$ 3889,3835,3797, and 3770 {\AA} are clearly visible in absorption and are
 blue-shifted relative to the Balmer emission lines by about 300 km  s$^{-1}$ which we attribute to formation in the expanding wind or ejecta. The Fe II multiplet 42 lines are also 
prominent in emission with strong P Cygni absorption features, but with no more than one 
absorption minimum, blue-shifted by about 400 km  s$^{-1}$ relative to the peak emission. So the Fe II absorption is not formed in the faster moving ejecta. 
Numerous  other Fe II  lines are  present in emission but without P Cygni profiles.  
The Ca II K absorption line is present, and in the red, the strong, luminosity sensitive  O I
triplet at $\lambda$ 7774 {\AA} is in absorption.

The first low-resolution spectrum described by \citet{Tartaglia} and observed within 24 hours
of the discovery, does not show the multiple P Cyg absorptions. Our first
spectrum was obtained only five days later, and although the difference 
could be real, we attribute it to the difference in spectral resolution, 
14 {\AA} vs. 3 {\AA} in our spectra. The total width of the P Cyg absorption 
feature is the same in both spectra.

All of the  line identifications and measured velocities discussed 
in this section 
are given in Table 4.  
Although,  it is common to quote the terminal velocity ($v_{\infty}$) for the P Cygni profiles of hot stars, the terminal velocity, however, is derived from  a stellar wind model fit to the profiles of resonance lines.  The Hydrogen lines are not resonance lines. We also chose  not to fit Gaussians to determine the blue edge velocity because  of the reduced S/N in some of the spectra together with the complex profiles with multiple minima plus the strong scattering wings. Instead, we give the velocity at the  absorption minimum measured relative to the emission peak which permits a well-controlled differential measurement.

When our second spectrum was observed on Day 71, PSN J09132750+7627410 had already faded significantly.
The blue spectrum has very poor S/N, and only H$\beta$ is identified in emission, however,  in 
the red, H$\alpha$ is still 
strongly in emission (Figure 7) with multiple  P Cygni absorption minima. The scattering 
wings have weakened considerably due to decreasing  density in the expanding ejecta. The red wing is now measured to only 900 km  s$^{-1}$.   In addition to the
O I triplet, the Ca II near-infrared triplet has appeared plus Fe II and Fe I  absorption lines and the K I doublet (Figure 6). The presence of these additional absorption lines suggests that the eruption
has produced the optically thick cool wind observed in several post-maximum giant eruptions 
such as SN2011ht \citep{RMH2012} and UGC 2773 OT2009-1 \citep{Smith,Foley}, the red transients SN2008S and the NGC 300 OT, as well as the 
 LBV/S Dor variables at maximum light. These absorption lines are offset by $\approx$ 400 km  s$^{-1}$ relative to the Hydrogen emission. 

\begin{figure}[h!]  
\figurenum{6}   
\epsscale{1.0}
\plotone{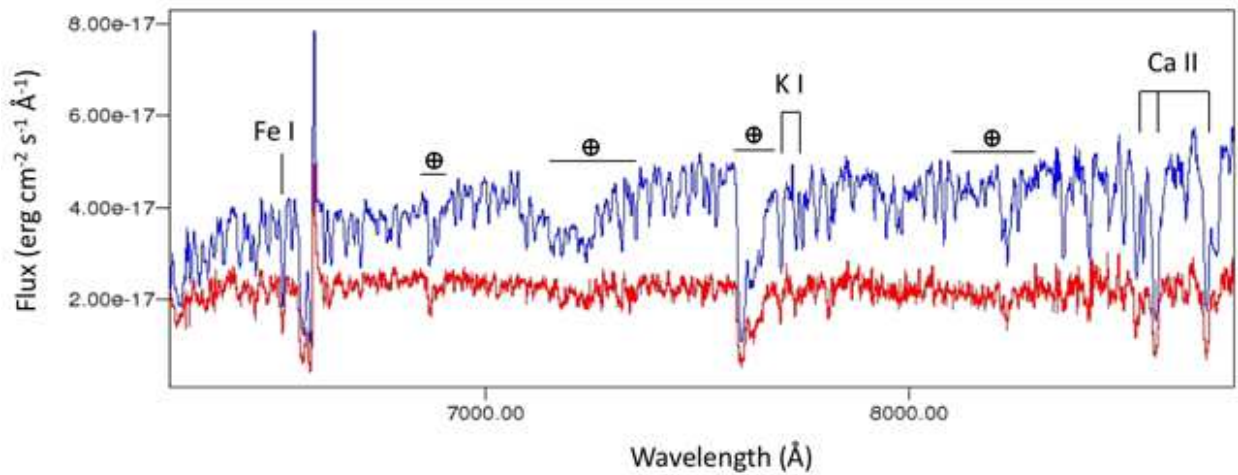}
\caption{The red spectra from Days  71 (red) and 99 (blue) illustrating the deve
lopment of the absorption lines. }
\end{figure}

Several additional absorption lines from neutral metals  appeared in both the blue and the red  
spectra from  Day 99. We note that the absorption lines present earlier show a 
redwards shift by about 50 - 100 km s$^{-1}$ between  Days 71 and 99.  The slow moving P Cyg absorption feature shows a similar redwards shift relative to 
the Hydrogen emission peak.  This could be due either to a slowing of the 
dense wind or to possible  infall back to the star. A slowing of the ejecta is often attributed to collision with previous circumstellar material but, in 
this case, the slow moving gas will have  reached only 20 AU in 99 days 
while 
the two faster winds will have expanded to 63 and 91 AU, and further if ejected earlier. The dust formation radius will depend on the properties of the star, but is typically at least 100 -- 200 AU and for the infrared source it is 
$\approx$ 100 AU. So what is it colliding with?

Despite a longer integration time the blue spectrum  has  relatively 
poor S/N, although H$\beta$ still shows absorption minima. The red spectrum is shown in Figure 6 together with the Day 71 spectrum. For comparison, the 
three H$\alpha$ profiles are shown together in Figure 7.  

\begin{figure}[h!] 
\figurenum{7}   
\epsscale{1.0}
\plotone{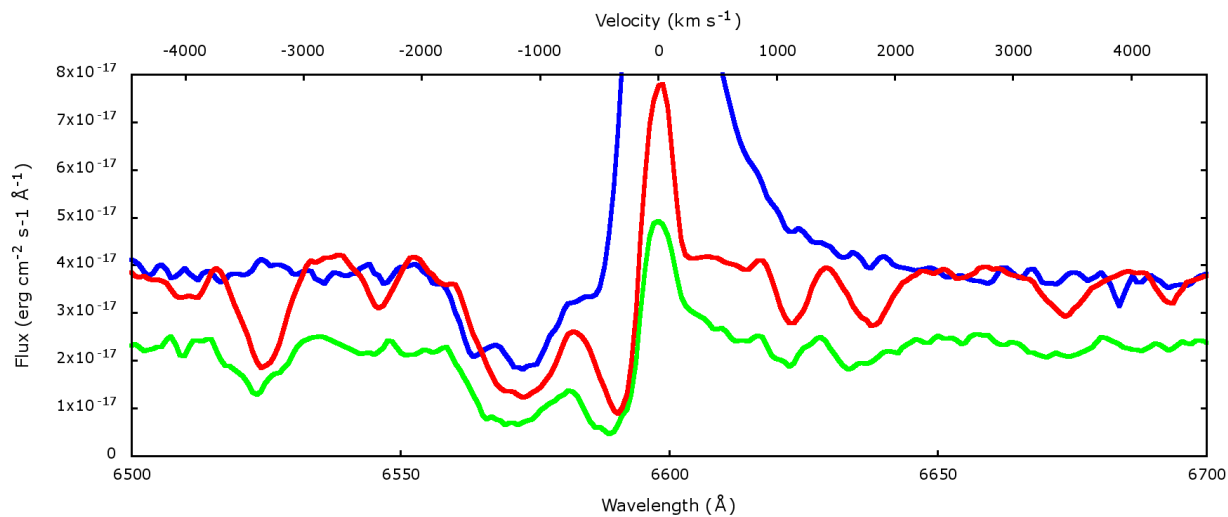}
\caption{The H$\alpha$ profiles from  all three spectra illustrating the P Cygni
 absorption shift to longer wavelengths with time. The top spectrum (blue)
  is from Day 6, green from Day 71 and red from  Day 99.}
\end{figure}

In general the spectra of PSN  J09132750+7627410 are typical of non-terminal giant eruptions 
with strong narrow Hydrogen emission, Thomson scattering profiles due to the strong wind, 
and expansion 
velocities of a few 100 km  s$^{-1}$. The distinguishing characteristic of our spectra of this 
eruption however, are the multiple P Cygni absorption features present in several of the
Balmer lines.  
The outflow velocities  measured  for the P Cygni absorption components from  our  three  
spectra are summarized in Table 5. 

Three distinct blue-shifted absorption features or minima are present at 
H$\beta$, H$\gamma$ and H$\delta$ in the spectra from Day 6 with blue-shifted velocities 
relative to their corresponding  emission peaks at $\sim$ -400, -1100 and -1600 km s$^{-1}$, hereafter respectively called velocity components {\it a}, {\it b}, and {\it c}, see Figure 5.  
The two highest velocity minima, {\it b} and {\it c},  are clearly present at H$\alpha$; 
component {\it a}, at the lowest velocity  
is very likely covered by the scattering wing. 
H$\epsilon$ shows a classical P Cygni profile, although component {\it c}  
 may be present as an unidentified absorption at 3968{\AA}. 
Unfortunately, only the red spectrum from Day 71 has sufficient signal to noise for 
reliable line measurements. The wings on the H$\alpha$ scattering profile have weakened considerably, and  
velocity component {\it a} is now clearly visible as a strong P Cygni feature. The two higher velocity minima, {\it b} and {\it c},  have weakened, and have merged into a single absorption feature although multiple components or structure  are 
visible in the profile.  
In our third spectrum from Day 99, the highest velocity component {\it c}  
is no longer recognizable at H$\alpha$. There are now two strong absorption features and the blue edge of the high velocity component has shifted noticeably 
redwards. Two absorption minima  corresponding to components {\it a} and 
{\it b} can 
still be clearly identified in the H$\beta$ P Cygni profile  and  component
{\it c} may still be present but much weaker.

The multiple absorption features or P Cygni profiles could be
due either to multiple outflows or ejections with different velocities and energies from
the current eruption or be a remnant of the wind and prior mass loss episodes of  the progenitor. If we assume
that the lowest velocity feature  at -400 km s$^{-1}$ is formed in the  expanding ejecta
in the current outburst, then there is clear evidence for two additional, 
separate winds or outflows. 
Their possible origin and implications for the eruption are discussed in the 
next section. 

\section{Discussion} 

The absorption minima at $\sim$ -400 km s$^{-1}$  plus the velocity difference
between the absorption lines and emission peaks of 300-400 km s$^{-1}$, and the FWHM of the narrow emission cores are all consistent with a relatively slow
ejection velocity 
which we associate with the current brightening or eruption and the formation of the cool, dense wind. Expansion 
velocities on this order are higher than measured for LBV/S Dor variables 
(100 -- 200 km s$^{-1}$) in their optically dense wind state or maximum light, 
but are observed in giant eruptions such as UGC 2773 OT1009-1 (350 km s$^{-1}$) and SN2009ip (550 km s$^{-1}$) \citep{Smith,Foley}, SN2011ht 500 -- 600 km s$^{-1}$ \citep{RMH2012} and even  $\eta$ Car (600 km s$^{-1}$),  although  
higher velocities have been reported in impostor eruptions such as SN2000ch (3000 km s$^{-1}$) \citep{Pastorello}.  
The outstanding question is the origin of the two higher velocity blue-shifted
absorption minima, {\it b} and {\it c},  in PSN J09132750+7627410.

We have identified four other objects in which multiple absorption minima 
have been observed; SN2005gj \citep{Trundle}, SN2009ip \citep{Margutti} and two giant eruptions,
SN2000ch \citep{Pastorello} and SNHunt 248 \citep{Mauerhan}. 
\citet{Trundle} measured velocities of 300 and 100 km s$^{-1}$ in two
absorption troughs in the H$\alpha$ and H$\gamma$ lines in SN2005gj which
they suggested were the remnants of the progenitor's previous mass loss state
as an LBV. In contrast, much higher velocities of 3000 and 5300 km s$^{-1}$
were  measured in  two absorption minima in the H$\beta$ line during 
SN2000ch's 2009-OT1 eruption \citep{Pastorello}, and similar minima appear to be present in the H$\gamma$ line. Although, \citet{Mauerhan} 
did not specifically discuss them, multiple absorption minima can be seen 
in their first spectrum of SNHunt 248
for which they suggest an average outflow of $\sim$ 1200 km s$^{-1}$.
In the peculiar SN2009ip multiple absorptions at high velocities are 
observed in the post-terminal eruption spectra, but were not reported in its 
prior eruptions when it was considered a non-terminal giant eruption. 

It is common in work on SN impostors and even terminal Type IIn SNe,
to assume that evidence for additional outflows in their spectra or mass loss 
episodes in the ejecta are
from separate preceding events. In several of these
objects, SN2011ht, SN2009ip, SN1994W, SN2009kn, etc., enhanced mass loss precedes the main
eruption by only a few months or years, as in SN2009ip.  Instead of 
being viewed as entirely
separate events occurring years or decades before, they could all be 
part of the same instability producing more than one eruption as part of the 
 on-going episode or instability. In the case of slow moving ejecta, it is often assumed that it is due to a prior LBV/S Dor stage
and that the progenitor was an LBV, but that might not be correct. In all of
these cases  we do not have a sufficient timeline to know. 

In PSN J09132750+7627410, the velocities at {\it b} and {\it c} are too high to 
be produced by the slow, dense wind of a presumed LBV progenitor or even  
LBVs in quiescence which have relatively slow winds for their corresponding
hot supergiant spectral types. 
If either  of these high velocity components are  
remnants of the progenitor's wind, then they imply a hot star.
One of these components could be the  remnant of the progenitor's wind and
the second from a previous, and relatively recent, high mass loss
episode.  These high wind  velocities
are measured in late O-type and early B-type stars \citep{Kud}, but neither star
A or B has an SED suggestive of hot star, and if the infrared source is the 
progenitor, the embedded star was most likely a cool or intermediate 
temperature supergiant.

Alternatively, both may be  from previous high mass loss episodes. 
Their much higher expansion velocities compared to the current event suggest 
much more energetic eruptions. Unfortunately there is no record of 
prior brightenings. However, the  infrared source was observed for five years
and was still present slightly more than a year prior to the eruption. If it was the
progenitor, this  may set  a limit on when the prior ejections occurred, and
an eruption, represented perhaps by one or both of these higher velocity outflows was responsible for the dust destruction. The dust condensation distance 
for the 780\arcdeg K infrared source is 100 AU. Velocity components {\it c} and
{\it b} would have reached this distance in a fraction of a year.

Here we suggest  that the  multiple P Cygni 
profiles were all formed during the current eruption  which may have 
begun several weeks or even months before the brightening was observed, and are 
due to material ejected at different velocities or energies and possibly at different times. $\eta$ Car
provides an example of ejecta expanding at a range of velocities from the 
same giant eruption; from 600 km s$^{-1}$ in the Homunculus lobes to
1000 -- 3000 km s$^{-1}$ in the outer ejecta (\citet{Weis} and references
therein). This possibility for PSN J09132750+7627410 is supported by the weakening and disappearance of the highest velocity component {\it c}  as the eruption subsided and the ejecta expanded over the 99 days covered by our spectra.

Formation during a single mass loss episode applies just as well to the two non-terminal objects described above. The multiple P Cyg minima in the first SNHunt 248 spectrum 
closely resemble the absorption troughs in our first spectrum of PSN J09132750+7627410, and  were not apparent in the later spectra observed over the next 
month. Although most had  lower resolution, one with higher resolution confirms that  P Cygni minima were  not present about one month after they were first observed.  A series of spectra by \citet{Kankare} shows a lack of multiple minima at the same time in their highest resolution spectrum. Thus, the multiple P Cyg absorptions quickly disappeared and could have a common origin as multiple ejections during the
high mass loss event.  The  velocities measured in SN2000ch  
 during its 2009-OT1 eruption seem high compared to these other objects,
but are comparable to the velocities reported in $\eta$ Car's outer ejecta. 
Like SNHunt 248, the P Cyg absorptions are gone in spectra obtained a few weeks later. SN2005gj is considered to be a terminal explosion, but that 
doesn't rule out the ejection of material shortly before the eruption. 
SN2009ip is an example where lesser ``giant eruptions'' occurred several times during the three years prior 
to what is assumed to be the final event. The multiple absorptions observed
in its post-maximum spectra with very high velocities of -5000 to -12000 
km s$^{-1}$, were produced during its final eruption and are considered  
evidence for separate asymmetric outflows \citep{Margutti}. Similar processes may be occurring in the less energetic
eruptions in the impostors. They did in $\eta$ Car. 

In the two non-terminal giant eruptions, the multiple absorption minima weaken 
and 
disappear within a few weeks. In PSN  J09132750+7627410 they were present for a couple of months. So it is possible that the evidence for multiple 
outflows during the high mass loss episodes is more common  than has been reported due to their relatively brief or transient appearance, and to    
lower spectral resolution, especially in the early post-discovery spectra, as we
found for  PSN J09132750+7627410.

We thank Kris Davidson for valuable  discussion and suggestions. Research by R. Humphreys and M. Gordon on massive stars is supported by  
the National Science Foundation AST-1109394. J. C . Martin's collaborative work 
on luminous variables is supported by the National Science Foundation grant  AST -1108890. Some of the data presented in this paper were obtained from the 
Mikulski Archive for Space Telescopes (MAST). STScI is operated by the 
Association of Universities for Research in Astronomy, Inc., under NASA 
contract NAS5-26555.   We also used data from  the Pan-STARRS Survey for 
Transients. Operation of the Pan-STARRS1 telescope is supported by the National Aeronautics and Space Administration under Grant No. NNX12AR65G and Grant No. NNX14AM74G issued through the NEO Observation Program.

{\it Facilities:} \facility{LBT/MODS1}


\begin{deluxetable}{lllll}
\tablewidth{0 pt}
\tabletypesize{\footnotesize}
\tablenum{1} 
\tablecaption{Journal of Spectroscopic Observations}
\tablehead{
\colhead{UT Date} &
\colhead{Instrument}  &
\colhead{Exp. Time} &
\colhead{Wavelength} &
\colhead{Reference} 
}
\startdata 
11.98  Feb 2015 (Day 1)  &  Asiago 1.8m &   \nodata  & 3400 -- 8200{$\AA$} & \citet{Tartaglia} \\ 
16.24 Feb 2015 (Day 6) &  LBT/MODS1   &   30m      & 3200{$\AA$} -- 1$\mu$m   &        \\
22.13 Apr 2015 (Day 71) &   "      &   30m      &    "                         &        \\ 
20.16  May   2015 (Day 99)   &   "      &   60m     &    "                         &        \\
\enddata
\end{deluxetable} 

\begin{deluxetable}{llllllll}
\rotate
\tabletypesize{\scriptsize}
\tablenum{2}
\tablecaption{Multi-color Photometry }
\tablehead{
\colhead{U.T. Date}  &
\colhead{B mag} &
\colhead{V mag} & 
\colhead{r(P1) mag} &
\colhead{R mag} &
\colhead{i(P1) mag} & 
\colhead{I mag}  &  
\colhead{Comment} 
}
\startdata
10.92 Feb 2015  &    \nodata  &    \nodata &    \nodata  &    \nodata & \nodata &     \nodata & 18.0 unfiltered; Cortini \\ 
11.15 Feb 2015  &  \nodata  &  \nodata &   \nodata  &    \nodata & \nodata &     \nodata &  17.7 unfiltered; Yusa \\ 
11.23 Feb 2015  &  \nodata  &  18.14 &  \nodata &    \nodata  &    \nodata &  17.42 &  Kiyota\\ 
11.43 Feb 2015  &    \nodata  &    \nodata &   \nodata  &    \nodata &  \nodata &     \nodata & 17.8 unfiltered; Noguchi\\
11.98 Feb 2015 &   \nodata  &  18.7  &    \nodata  &    \nodata &   \nodata      &  \nodata   &  \citet{Tartaglia} \\
26.91 Feb 2015 &  \nodata  &    \nodata &  18.34 $\pm$ 0.03 & \nodata  &   \nodata &   \nodata & Pan-STARRS\\
12.66 Mar 2015 &   18.83 $\pm$ 0.08 & 18.58 $\pm$ 0.12 & \nodata &  17.54 $\pm$ 0.05 & \nodata &   \nodata & Barber Obs.\\
17.63 Mar 2015 &   18.69 $\pm$ 0.11 & 18.22 $\pm$ 0.12  &  \nodata & 17.58 $\pm$ 0.05 & \nodata & 17.40 $\pm$ 0.09  & Barber Obs.\\
28.76 Mar 2015 &    \nodata  &    \nodata & \nodata  &    \nodata & 18.80 $\pm$ 0.02  & \nodata & Pan-STARRS\\ 
03.82 Apr 2015 &    \nodata  &    \nodata & \nodata  &    \nodata & 18.94 $\pm$ 0.03 & \nodata & Pan-STARRS\\
29.60  Apr 1015 &   $>$18.8   &     $>$ 18.9   &  \nodata &  $>$18.5   & \nodata & \nodata & Barber Obs\\
19.67 May 2015  &   $>$ 19.0   &     $>$ 19.3   &  \nodata &  18.22 $\pm$ 0.09  &  \nodata & \nodata & Barber Obs\\
13.73 Jan 2016  &  $>$ 19.2   &   $>$19.0   &   \nodata &  $>$18.6 &  \nodata & \nodata & Barber Obs\\
5.15 Feb 2016  &   $>$19.5   &   $>$ 19.2 &   \nodata &  $>$18.8  & \nodata & \nodata & Barber Obs\\
\enddata
\end{deluxetable} 

\begin{deluxetable}{lllllll}
\tabletypesize{\footnotesize}
\tablenum{3}
\tablecaption{HST/WFPC2 and Spitzer/IRAC Photometry}
\tablehead{
\colhead{Date}  & \colhead{Star}  & \colhead{F450W mag.}   & \colhead{F814W mag.} &  \colhead{3.6$\mu$m mag.} & \colhead{4.5$\mu$m mag.}  &  \colhead{ICRS J2000 Position} 
}
\startdata  
6 Jul 2001 &    A    &  24.43 $\pm$ 0.20 & 23.79 $\pm$ 0.15 &  ...   &  ... & 
9:13:27.48 +76:27:41.1\\
 "    &    B &   26.08	$\pm$ 	1.28 & 	24.00$\pm$ 0.18 &  ...   &  ...  &       9:13:27.38 +76:27:40.9 \\
2 Dec 2009 &  ... & ... & ... &  19.41 $\pm$ 0.3 &  18.40 $\pm$ 0.3 &  \nodata \\
12 Apr 2010  &  ... & ... & ... & 19.16 $\pm$ 0.2 & 18.65 $\pm$ 0.3 &  \nodata \\
25 Jan 2014  &  ... & ... & ... &  19.21 $\pm$ 0.2 &  18.32 $\pm$ 0.1 &  9:13:27.24 +76:27:41.2  \\
\enddata
\end{deluxetable}

\begin{deluxetable}{lllll}
\rotate
\tabletypesize{\scriptsize}
\tablenum{4}
\tablecaption{Line Identifications and Measured Velocities in PSN J09132750+7627410}
\tablehead{
\colhead{Line}  &
\colhead{Velocity km s$^{-1}$}  &
\colhead{P Cyg Vel. km s$^{-1}$}  &
\colhead{Blue-Edge Vel. km s$^{-1}$}  &
\colhead{Red Wing km s$^{-1}$}   
}
\startdata 
         &            &  Day 6 (16 Feb 2015) &   &     \\
Emission Lines &      &       &     &         \\
H$\alpha$  & 1607  &  \nodata, -1137, -1537  &  \nodata, -1337,  -1817 &  2460   \\
H$\beta$  &  1648  & -466, -1154, -1559      &  -681,  -1353, -1946   &  2200    \\
H$\gamma$ &  1680  & -412, -1027, -1622       &  -646,  -1127, -1876   &  1650   \\
H$\delta$ &  1745  & -426, -1124, -1670      & -717, -1255,  -1953     &  1668    \\
H$\epsilon$ & 1745  & -428, \nodata, \nodata\tablenotemark{a}  &  \nodata, \nodata, \nodata    \\
Fe II(42) 4923.9{\AA} &  1682  &   -409   & \nodata  &  \nodata  \\
Fe II(42) 5018.4{\AA} &  1680  &   -348   & \nodata   &  \nodata   \\
Fe II(42) 5169.0{\AA} &  1706  &   -427   & \nodata   &  \nodata   \\
Absorption lines  &      &       &     &         \\ 
H 3889{\AA}  &  1365 & \nodata   &  \nodata    &  \nodata        \\ 
H 3835{\AA}  &  1408 & \nodata   &  \nodata    &  \nodata        \\
H 3797{\AA}  &  1335 & \nodata   &  \nodata    &  \nodata        \\
H 3770{\AA}  &  1313 & \nodata   &  \nodata    &  \nodata        \\
Ca II K      &  1391 & \nodata   &  \nodata    &  \nodata        \\
O I 7774{\AA} & 1397 & \nodata   &  \nodata    &  \nodata        \\
         &            & Day 71 (22 Apr 2015) &   &     \\
Emission Lines &      &       &     &         \\
H$\alpha$ & 1595  &  -396, -1171, -1453:: &  -727, -1764  &  914    \\
Absorption lines  &      &       &     &         \\
O I 7774{\AA} &  1389   & \nodata   &  \nodata    &  \nodata        \\
Ca II8498{\AA} & 1239   & \nodata   &  \nodata    &  \nodata        \\
Ca II8542{\AA} & 1296   & \nodata   &  \nodata    &  \nodata        \\
Ca II8662{\AA} & 1306   & \nodata   &  \nodata    &  \nodata        \\
K I 7665{\AA}  &  1211  & \nodata   &  \nodata    &  \nodata        \\
K I 7699{\AA}  & 1230   & \nodata   &  \nodata    &  \nodata        \\
Fe II 6432{\AA} & 1240  & \nodata   &  \nodata    &  \nodata        \\
Fe I  6494{\AA} & 1289  & \nodata   &  \nodata    &  \nodata        \\
        &            &  Day 99 (20 May 2015)  &   &     \\
Emission Lines &      &       &     &         \\
H$\alpha$ &  1614     &  -364, -1161, \nodata  & -682, -1735  &  909   \\
H$\beta$  &  1651     &  -316, -913, range   &  -636, -1047,   & \nodata   \\
H$\gamma$ & 1701     & \nodata   &  \nodata    &  \nodata        \\
Absorption lines  &      &       &     &         \\
O I 7774{\AA} &  1420  & \nodata   &  \nodata    &  \nodata        \\
Ca II8498{\AA} & 1318  & \nodata   &  \nodata    &  \nodata        \\
Ca II8542{\AA} & 1333  & \nodata   &  \nodata    &  \nodata        \\
Ca II8662{\AA} & 1340  & \nodata   &  \nodata    &  \nodata        \\
K I 7665{\AA}  &  1295 & \nodata   &  \nodata    &  \nodata        \\
K I 7699{\AA}  &  1341 & \nodata   &  \nodata    &  \nodata        \\
Fe II 6432{\AA} &  1288 & \nodata   &  \nodata    &  \nodata       \\
Fe I  6494{\AA} &  1366 & \nodata   &  \nodata    &  \nodata       \\
Fe I  8327{\AA} &  1329 & \nodata   &  \nodata    &  \nodata       \\
Fe I  8387{\AA} &  1283 & \nodata   &  \nodata    &  \nodata        \\
Na I  8194{\AA} &  1333 & \nodata   &  \nodata    &  \nodata        \\
\enddata
\tablenotetext{a}{An unidentified line at 3968{\AA} may be the blue-shifted 
component c with a corresponding velocity of -1773 km s$^{-1}$.} 
\end{deluxetable}

\begin{deluxetable}{lll}
\tabletypesize{\footnotesize}
\tablenum{5}
\tablecaption{Summary of the Outflow Velocities}
\tablehead{  
\colhead{Date}  &
\colhead{Velocity km s$^{-1}$}  &
\colhead{Lines}   
}
\startdata
Day 6     &    -433     &  H$\beta$, H$\gamma$, H$\delta$, H$\epsilon$   \\
          &    -1110    &  H$\alpha$, H$\beta$, H$\gamma$, H$\delta$    \\
	  &    -1597    &  H$\alpha$, H$\beta$, H$\gamma$, H$\delta$   \\
	  &    -395     &  Fe II (42)        \\
          &             &                    \\
Day 71  &   -396   &  H$\alpha$     \\
        &   -1171  &  H$\alpha$     \\ 
	&   -1453  &  H$\alpha$     \\ 
	&             &                                     \\
Day 99  &   -340   &  H$\alpha$, H$\beta$                   \\
        &  -1161, -913  &  H$\alpha$, H$\beta$                   \\
\enddata
\end{deluxetable}

\end{document}